\documentclass[12pt]{article}
\usepackage{amsmath}
\usepackage{amssymb}
\numberwithin{equation}{section}

\begin{document}
\baselineskip=17pt
\begin{titlepage}
\begin{flushright}
{\small KUNS-2155}
\end{flushright}
\begin{center}
\vspace*{8mm}

\begin{minipage}{1.05\linewidth}{\large\bf%
$\!\!$Flavor Symmetry Breaking and Vacuum Alignment on Orbifolds}
\end{minipage}
\vspace*{4mm}

Tatsuo Kobayashi, \ Yuji Omura, \ Koichi Yoshioka
\vspace*{3mm}

{\it Department of Physics, Kyoto University,
Kyoto 606-8502, Japan}
\vspace*{3mm}

{\small (September, 2008)}
\end{center}
\vspace*{5mm}

\begin{abstract}\noindent%
Flavor symmetry has been widely studied for figuring out the masses
and mixing angles of standard-model fermions. In this paper we present
a framework for handling flavor symmetry breaking where the symmetry
breaking is triggered by boundary conditions of scalar fields in
extra-dimensional space. The alignment of scalar expectation values is
achieved without referring to any details of scalar potential and its
minimization procedure. As applications to non-abelian discrete flavor
symmetries, illustrative lepton mass models are constructed where 
the $S_3$ and $A_4$ flavor symmetries are broken down to the
directions leading to the tri-bimaximal form of lepton mixing and
realistic mass patterns.
\end{abstract}
\end{titlepage}

\section{Introduction}

The masses and mixing angles of quarks and leptons have been
long-standing and inspiring problems in particle physics. Their
observed patterns, e.g.\ the recent experimental result for neutrino
generation mixing~\cite{NeuExp}, suggest underlying principles such as
grand unification and flavor symmetry lying behind the data.

Flavor symmetry has been widely investigated as a solution to the
fermion mass problems. Among various flavor groups, a particular
attention has recently been paid for non-abelian discrete groups such
as $S_3$ and $A_4$\@. These non-abelian discrete symmetries have
several advantages that they provide a definite meaning for the
generation and also link up different generations. However flavor
symmetry should be broken in a non-trivial way while preserving the
trace of symmetry structure. Flavor symmetry breaking is caused, e.g.,
by non-vanishing vacuum expectation values of scalar fields, which are
determined by the minimization of scalar potential. For a non-abelian
discrete symmetry to leave some imprints in low-energy symmetry-broken
theory, the expectation values must take specific directions, namely,
the vacuum alignment is required. That is usually achieved by
performing the detailed analysis of scalar potential invariant under
flavor symmetries~\cite{align,align2}. In particular, the vacuum
alignment has recently been discussed in numbers of lepton mass models
which account for the tri-bimaximal form of generation
mixing~\cite{TBM} as a good approximation of the experimental
data. The tri-bimaximal mixing implies a specific form of mass matrix
for light Majorana neutrinos: $M_L=\sum_im_iv_iv_i^{\rm t}$ with 
$v_1=(2,-1,-1)$, $v_2=(1,1,1)$ and $v_3=(0,1,-1)$. These inter-family
related values of matrix elements seem to indicate non-abelian
symmetry, vacuum alignment, and so hidden structure of nature beyond
the standard model.

In this paper, we present a framework to realize aligned flavor
symmetry breaking without any elaborated potential analysis. A central
ingredient is the global property of bulk scalar fields in
higher-dimensional space. At the boundaries of extra space, dynamical
conditions for bulk fields are specified to fix the model. That
reduces the number of degrees of freedom of symmetry 
groups~\cite{SS}. This type of symmetry breaking with extra dimensions
has been applied to, for example, the electroweak gauge
symmetry~\cite{EW_break} and grand unified theory~\cite{GUT_break}. As
for flavor symmetry, a viable breaking mechanism is provided by
flavor-twisted boundary conditions for bulk fermion
fields~\cite{HWY}. In this paper we discuss bulk scalar fields charged
under the flavor symmetry, whose expectation values are aligned to
cause discrete patterns of symmetry breaking. The vacuum alignment is
archived in our approach by adopting the boundary conditions such
that only one light mode survives and it governs all vacuum expectation
values in low-energy theory. Our method is also applicable to the
quark sector and grand unified theory.

This paper is organized as follows. In Section 2, we first present the
general procedure to have aligned scalar expectation values and
classify the flavor-group matrix elements in the five-dimensional
orbifold theory. The next two parts (Sections 3 and 4) are devoted to 
the applications of $S_3$ and $A_4$ flavor symmetries. We also
present illustrative lepton mass models with flavor symmetries and
vacuum alignments which are implemented on higher-dimensional
orbifolds. Section 5 summarizes our results.

\bigskip
\section{General recipe}

In this section, we consider as the simplest example a
five-dimensional orbifold theory on the flat gravitational
background. The generalization to higher dimensions is
straightforward. The five-dimensional coordinates are denoted 
by $(x^\mu,x^5)$ where $x^5$ corresponds to the fifth dimension. There
are two types of operations on this space; the 
reflection $\hat Z$~:~$x^5\to-x^5$ and the 
translation $\hat T$~:~$x^5\to x^5+2\pi L$, where $L$ is a
constant. They trivially satisfy $\hat T\hat Z=\hat Z\hat T^{-1}$. On
field variables living in the bulk, these operations are expressed in
terms of matrices in the field space:
\begin{eqnarray}
  \hat Z\phi(x^5) \,=\, Z^{-1}\phi(-x^5), \qquad
  \hat T\phi(x^5) \,=\, T^{-1}\phi(x^5+2\pi L).
\end{eqnarray}
Let us suppose the extra space is compactified on 
the $S^1/Z_2$ orbifold with the radius $L$. This is achieved by the
identifications $\hat Z(x^5)=x^5$ and $\hat T(x^5)=x^5$. On this
orbifold, there are two fixed 
points (boundaries) at $x^5=0$ and $x^5=\pi L$. The boundary 
conditions of bulk fields in compactifying the extra dimension are
then defined by the identifications $\hat Z\phi(x^5)=
\phi(x^5)$ and $\hat T\phi(x^5)=\phi(x^5)$.

It is noted that possible boundary conditions for bulk fields are
limited by several consistency relations. First, 
since $\hat Z$ behaves as parity, available choices are
\begin{equation}
  Z^2 \,=\, 1.
\end{equation}
In addition, for the translation, the matrix $T$ must satisfy the 
relation
\begin{equation}
  ZT \,=\, T^{-1}Z.
  \label{consistency}
\end{equation}
Clearly not all types of boundary conditions are allowed. For example,
in the case that $Z$ is trivial (the unit matrix), the translation
matrix $T$ is limited to $T^2=1$. That is more easily seen by defining
another parity operation $\hat Z'=\hat T\hat Z$, which is interpreted
as the reflection about $x^5=\pi L$. The consistency 
relation \eqref{consistency} just implies $(Z')^2=1$, where $Z'$ is
the matrix representation of $\hat Z'$ on the field space.

With a set of boundary conditions, bulk fields are expanded by
Kaluza-Klein modes. After integrating over the fifth dimension, we
obtain a four-dimensional effective theory below the compactification
scale $1/L$. In the low-energy theory, the physics is described by
zero modes. In this paper, we focus on bulk scalar fields. They are
assumed to be non-trivially charged under the flavor symmetry. The
boundary condition determines which components of bulk scalars 
have (massless) zero modes. These zero modes are expected to develop
their vacuum expectation values in low-energy effective theory and
then contribute to Yukawa operators for quarks and leptons. The
expectation values are determined, for example, \`a la
Coleman-Weinberg~\cite{CW} or by use of boundary terms whose
existence is however irrelevant to the following discussion, and
further details are not considered in this paper. On the other hand,
other excited modes are superheavy and stabilized at the origin with
vanishing expectation values.

An idea is that a specific boundary condition is adopted under which
only one zero mode remains in low-energy theory. All non-vanishing
expectation values are controlled by this single zero mode and
therefore related to each other, i.e., the vacuum alignment is
established. If there are more than one zero modes, several directions
of the vacua are degenerate and generally depend on free model
parameters. It is noted that this procedure makes sense in
phenomenology only for non-abelian and discrete symmetries. For
an abelian or continuous symmetry, the vacuum described by multi field
values is generally determined by continuous parameters and therefore
the alignment does not work.

Here we give a general recipe for non-abelian discrete symmetry
breaking in the fifth dimension $S^1/Z_2$ such that only a single zero
mode survives. The bulk theory has a flavor symmetry acting on the
three-generation quarks and leptons. We introduce bulk scalar 
fields $\phi_i$ ($i=1,2,3$) which transform as a triplet under the
flavor symmetry. The triplet representation can be reducible. The
boundary conditions for the scalars $\phi_i$ are given by
\begin{alignat}{3}
  \hat Z &:& \qquad \phi(-x^5) &\,=\, Z\phi(x^5), \\
  \hat T &:& \qquad \phi(x^5+2\pi L) &\,=\, T\phi(x^5).
  \label{Tbc}
\end{alignat}
Here $Z$ and $T$ are the unitary representation matrices of flavor
symmetry group and act on the generation space. The consistency
relations require $Z^2=(ZT)^2=1$. This implies that for each choice of
the matrix $Z$, it can be rotated to the diagonal 
form; $Z=\text{diag}(p_1,p_2,p_3)$ with $p_i=\pm1$. For a 
positive (negative) parity element $p_i=+1$ ($p_i=-1$), a
corresponding bulk field has an even (odd) wavefunction about the
extra-dimensional coordinate. Therefore the number of zero modes is
given by that of positive parity elements when one takes into
account the $\hat Z$ boundary condition only. A fast conclusion at this
stage is that at least one positive parity elements are needed to have
one zero mode. Therefore we fix $p_1=+1$ in the following discussion
in this section. To include another boundary condition taken into
account, it is useful to express the translation matrix 
as $T=e^{\pi iW}$ where $W$ is the hermite $3\times3$ matrix written
in the basis where $Z$ is diagonal. From the boundary 
condition \eqref{Tbc}, the mode expansion is found to be 
\begin{eqnarray}
  \phi_i(x^\mu,x^5) \,=\, \sum_n 
  \Big(\exp\Big[i\Big(\frac{W}{2}+n\Big)\frac{x^5}{L}\Big]\Big)_{ij}
  \phi^{(n)}_j(x^\mu).
  \label{KKmode}
\end{eqnarray}
The even (cosine) or odd (sine) function part is chosen depending on
the parity assignment for $\phi_i$. From this mode expansion, we find
that massless zero modes are obtained when $W$ has zero 
eigenvalues (mod 2). There are two types of boundary conditions for
realizing a single zero mode:
\begin{itemize}
\setlength{\itemsep}{7pt}
\setlength{\parskip}{0pt}
\setlength{\parindent}{15pt}
\item[I.] The number of massless mode is reduced to 1 with 
either $Z$ or $T$ only. This is the case that the matrix $Z$ has one
positive parity, i.e., $p_1=+1$ and $p_2=p_3=-1$. In this 
basis, $W$ should take the following form:
\begin{eqnarray}
W \,=\,
\begin{pmatrix}
0 & 0 & 0 \\
0 & * & * \\
0 & * & * \\
\end{pmatrix}.
\label{WT1}
\end{eqnarray}
The right-bottom $2\times2$ sub-matrix (denoted hereafter as $w$) is
hermite and has integer eigenvalues so that the consistency 
relation $(ZT)^2=1$ is satisfied. The translation $\hat T$ is
commutative with $\hat Z$ and becomes a parity operation, but not 
necessarily flavor diagonal. The relative rotation 
between $Z$ and $W$ is undetermined at this stage, but such a freedom
is irrelevant to the low-energy phenomenology of zero modes. We thus
find that a single zero mode appears in $\phi_1$. For the flat extra
dimension, the roles of two boundaries at $x^5=0$ and $x^5=\pi L$ are
interchangeable. This exchange gives another solution, 
namely $Z'=\text{diag}(+1,-1,-1)$ and $w$ with integer
eigenvalues. For any flavor symmetry group, we have such group
elements of the form \eqref{WT1}. For example, $T=1$ for $w=0$ 
and $T=Z$ for $w=\text{diag}(\pm1,\pm1)$.

\item[II.]
Another type of boundary conditions is that $Z$ and $T$ are in
conspiracy to reduce the number of zero modes. This is the case that
the matrix $Z$ has two positive parities, 
i.e., $p_1=p_2=+1$ and $p_3=-1$. The corresponding $W$ has either of
the following two forms.

The first case is that, like the Type I boundary condition, the
translation matrix $T$ becomes a parity and commutes with $Z$. It
takes the form
\begin{eqnarray}
W \,=\,
\begin{pmatrix}
* & * & 0 \\
* & * & 0 \\
0 & 0 & *
\end{pmatrix}.
\label{WT2}
\end{eqnarray}
The 3-3 element in the matrix \eqref{WT2} is an integer, but the even
integer case is equivalent to that in Type I if suitable exchanges of
the generation indices and two boundaries are performed. The
consistency relation $(ZT)^2=1$ and the requirement of a surviving
zero mode restrict the left-top sub-matrix has one even and one odd
eigenvalues. The zero mode is found in a definite linear combination
of $\phi_1$ and $\phi_2$. If the matrix $W$ is expressed in a basis
where the first component has a zero mode, it has the 
form $W=\text{diag}(0,1,1)$.

The second case is that $W$ is non-commutative with $Z$. We find the
most general expression to satisfy $(ZT)^2=1$ and to have a single
zero mode:
\begin{eqnarray}
W \,=\,
\begin{pmatrix}
& & \!\beta^* \\
& & \!\alpha^* \\
\beta & \alpha &  
\end{pmatrix},
\label{WT3}
\end{eqnarray}
with $\alpha$ and $\beta$ being arbitrary complex numbers. The
exception is that $|\alpha|^2+|\beta|^2=(\text{even integer})^2$ in
which case there appear two zero modes. The translation 
matrix $T$ obtained from \eqref{WT3} coincides with \eqref{WT2} just 
on the point $\alpha=1$ and $\beta=0$ (or vice versa), at which we
have a zero mode in $\phi_1$.

Several comments are in order. The interchange of $Z$ and $Z'$ also
gives a solution. Unlike the Type I boundary condition, there is no
relative freedom between $Z$ and $T$. This fact indicates the
conspiracy of flavor symmetry breaking. Note also that the Type II
boundary condition is viable only when the flavor symmetry group has a
matrix representation $X$ with $\det X=-1$.
\end{itemize}

\medskip

To summarize the result, the general recipe for obtaining a single
zero mode is the following. First, we diagonalize all representation
matrices for a triplet and list up the parity matrices with one
positive and two negative eigenvalues. The boundary conditions with
this class of elements directly lead to one zero mode and realize the
Type I boundary condition. The parity matrices with two positive and
one negative eigenvalues are also useful if combined with the specific
forms of translation, i.e.\ \eqref{WT2} or \eqref{WT3}. The 
matrix \eqref{WT2} has one even and one odd integer eigenvalues in the
left-top sub-matrix. The $Z$ and $T$ operations conspire to lead to a
single zero mode, called the Type II boundary condition. Through these
procedures, one can list up all possible boundary conditions. Finally,
we write down such boundary conditions in a unified basis.

This recipe is easily extended to higher-dimensional theories, for
example, the six-dimensional theory on the 
orbifold $T^2/Z_N$ for $N=2,3,4,6$.\footnote{See for various types of
orbifolds in higher dimensions, e.g.~\cite{orbifold}.} The only
difference from the present case is the form of consistency relations
and mode expansion.

\bigskip
\section{$\boldsymbol{S_3}$ flavor symmetry}

We first consider the permutation group $S_3$ as an example of flavor
symmetry, which is the simplest non-abelian discrete group and has
attractive features for flavor phenomenology~\cite{S3}. The $S_3$
group is composed of six elements which are the identity $I$, two
types of cyclic rotations $R$, $R^2$, and three types of odd
permutations $P$, $PR$ and $RP$. They satisfy the characteristic
relations for the $S_3$ group:
\begin{eqnarray}
  P^2 \,=\, (PR)^2 \,=\, R^3 \,=\, 1.
\end{eqnarray}
In addition to the trivial singlet, the $S_3$ group has two
non-trivial representations, the doublet and pseudo singlet, the
latter of which changes the sign under a 
permutation. Consequently, $S_3$ has two types of (reducible) triplet
representations: the triplet $3_S$ and the pseudo triplet $3_A$. On 
the field space, the circulation matrices are commonly represented on
both triplets but the permutation matrices are different. They are
explicitly written in the three-generation space by
\begin{eqnarray}
  P_A =-\begin{pmatrix}
  \,1 & & \\ & & \,1 \\ & \,1\, & \end{pmatrix}, \quad
  P_S =\frac{1}{3}\!\begin{pmatrix}
  -1\! & 2 & 2 \\ 2 & 2 & \!-1 \\ 2 & \!\!-1\!\! & 2 \end{pmatrix},
  \quad
  R =\begin{pmatrix} & \,1\, & \\ & & \,1 \\ \,1 & & \end{pmatrix},
\end{eqnarray}
and their derivatives $R^2$, $P_xR$ and $RP_x$ ($x=A,S$). The parity
operations $\hat Z$, $\hat Z'$ are therefore given 
by $P_x$, $P_xR$, and $RP_x$. Their representation matrices have the
determinants $+1$ and $-1$ for $3_A$ and $3_S$, respectively.

\medskip

\subsection{Vacuum alignment}

For a scalar $\phi_A$ in the $3_A$ representation, only Type I
boundary condition is possible since the parity matrices on $3_A$ have
two negative eigenvalues. Let us consider the $\hat Z$ boundary
condition associated with the twisting by $P_A$, i.e.,
\begin{eqnarray}
  \phi_A(-x^5) \,=\, P_A\,\phi_A(x^5).
\end{eqnarray}
The boundary condition with $P_AR$ or $RP_A$ is equivalent to that
with $P_A$ just by label exchange. The matrix $P_A$ is diagonalized by
the unitary transformation $\phi_A = U_A\bar\phi_A$:
\begin{eqnarray}
  U_A^\dagger P_A U_A \,=\, \begin{pmatrix}
  1\! & & \\ & \!\!-1\!\! & \\ && \!-1
  \end{pmatrix}, \qquad
  U_A \,=\, \begin{pmatrix}
  0 & \frac{1}{\sqrt{3}} & \frac{-2}{\sqrt{6}} \\
  \frac{-1}{\sqrt{2}} & \frac{1}{\sqrt{3}} & \frac{1}{\sqrt{6}} \\
  \frac{1}{\sqrt{2}} & \frac{1}{\sqrt{3}} & \frac{1}{\sqrt{6}}
  \end{pmatrix}.
  \label{diagPA} 
\end{eqnarray}
The diagonalizing matrix $U_A$ has a freedom in that its 
left-bottom $2\times2$ sub-matrix is undetermined due to the
degenerate eigenvalues. However that is irrelevant to the following 
discussion of zero mode physics. From \eqref{WT1} (i.e.\ the
consistency relation), the boundary condition with the 
translation $\hat T$ is found to be given by the 
matrix $T=I$ or $P_A$. Either choice for $T$ leads to the same
zero-mode physics. It is found from the parity sign 
in \eqref{diagPA} that only $\bar\phi_A^{\,1}$ has the massless zero
mode with a constant wave function. It survives in four-dimensional
low-energy effective theory and is expected to develop a vacuum
expectation value. In terms of the original fields, the vacuum
expectation values become
\begin{eqnarray}
  \langle\phi_A\rangle \,=\, \langle U_A\bar\phi_A\rangle
  \,\propto\, (0,-1,\,1).
\end{eqnarray}
Thus the vacuum described by $\phi_A$ is determined discretely, that
is, the vacuum alignment is achieved.

For a scalar $\phi_S$ in the $3_S$ representation, the Type II
boundary condition is viable. Similar to the $\phi_A$ case, 
the $\hat Z$ boundary condition is given by
\begin{eqnarray}
  \phi_S(-x^5) \,=\, P_S\,\phi_S(x^5).
\end{eqnarray}
The matrix $P_S$ is diagonalized as follows ($\phi_S=U_S\bar\phi_S$):
\begin{gather}
  U_S^\dagger P_S U_S \,=\, \begin{pmatrix}
  \,1\! & & \\ & 1\!\! & \\ && \!-1
  \end{pmatrix}, \qquad 
  U_S \,=\, \begin{pmatrix}
  \frac{1}{\sqrt{3}} & \frac{1}{\sqrt{2}} & \frac{1}{\sqrt{6}} \\
  \frac{1}{\sqrt{3}} & \frac{-1}{\sqrt{2}} & \frac{1}{\sqrt{6}} \\
  \frac{1}{\sqrt{3}} & 0 & \frac{-2}{\sqrt{6}} \\
  \end{pmatrix}.
  \label{diagPS}
\end{gather}
The parity signs 
in \eqref{diagPS} imply $\bar\phi_S^{\,1}$ and $\bar\phi_S^{\,2}$
could have massless zero modes. The Type II boundary condition further
needs non-trivial $\hat T$ twisting \eqref{WT2} or \eqref{WT3},
satisfying the consistency relation. It is easily found that, in the 
basis \eqref{diagPS} where $P_S$ is diagonal, no $S_3$ group element
satisfies \eqref{WT2}, but the matrix $R$ (and $R^2$) has the 
form \eqref{WT3} with $\alpha=-2i/3$ ($\alpha=+2i/3$). Therefore,
imposing the $\hat T$ boundary condition
\begin{eqnarray}
  \phi_S(x^5+2\pi L) \,=\, R\,\phi_S(x^5),
\end{eqnarray}
we find the mode expansion \eqref{KKmode} with the translation matrix
of the form \eqref{WT3}, and then obtain a single zero mode 
in $\bar\phi_S^{\,1}$. The vacuum expectation values of $\phi_S$ become
\begin{eqnarray}
  \langle\phi_S\rangle \,=\, \langle U_S\bar\phi_S\rangle
  \,\propto\, (1,\,1,\,1).
\end{eqnarray}
Thus the vacuum described by $\phi_S$ is aligned to the specific
direction. Notice that this direction is independent and physically
different from that of $\phi_A$. They are not converted to each other
by label exchanges.

\medskip
\subsection{Illustrative model}

With the aligned vacuum at hand, it is an interesting task to
construct realistic flavor symmetric theory for fermion masses and
mixing in higher dimensions. In this section, we present an
illustrative neutrino model based on the $S_3$ flavor symmetry.

Let us consider a six-dimensional model on the 
orbifold $T^2/(Z_2)^2$. While a straightforward application of the
previous result leads to a viable five-dimensional model with the
vacuum alignment, we here take a six-dimensional extension for Yukawa
operators and flavor symmetry invariance (see the last part of this
subsection). All the standard-model fields including the Higgs 
doublet $h$ and three-generation lepton 
doublets $\ell_i$ ($i=1,2,3$) as well as the right-handed 
neutrinos $\nu_i$ are confined on the four-dimensional fixed point 
at $x^5=x^6=0$. Besides the gravity, only standard-model gauge singlet
fields can propagate in the extra space not to violate the charge
conservations. We thus introduce gauge singlet five-dimensional
scalars $\phi_A$ and $\phi_S$, which induce effective neutrino Yukawa
operators. The five-dimensional fields $\phi_A$ and $\phi_S$ are
extended to the fifth and sixth dimension, respectively.

The bulk theory has the flavor $S_3$ and parity symmetries (other than
the parity operation~$\hat Z$ in the extra dimensions) under which the
fields have the following charges:\footnote{~The right-handed 
neutrino $\nu_1$ may belong to $1_S$ if it has a large Majorana mass
and is decoupled.}
\begin{eqnarray}
\begin{array}{c|ccccc|cc}
 & \ell & \nu_1 & \nu_2 & \nu_3 & h & \phi_S & \phi_A \\ \hline
S_3 & 3_S & 1_A & 1_S & 1_S & 1_S & 3_S & 3_A \\
Z  & + & + & - & + & + & - & + \\
Z' & + & + & + & - & + & + & -
\end{array}
\end{eqnarray}
In addition to the kinetic and potential terms for each field, we have
boundary interactions between the bulk and boundary fields, which give
the effective Yukawa operators invariant under the flavor symmetry:
\begin{eqnarray}
  {\cal L}_Y \,=\, 
  \bar\ell_i\big(y_S^{}\delta_{ij}
  +y_S'D_{ij}\big){\phi_S}_j\nu_2h +
  \bar\ell_i\big(y_A^{}\delta_{ij}
  +y_A'D_{ij}\big){\phi_A}_j\nu_3h + \cdots
  \label{LY_S3}
\end{eqnarray}
where the ellipsis denotes higher-dimensional operators 
including $\nu_1$ and the bulk field wavefunctions are evaluated at
the fixed point. These terms are described by two types of
symmetry-invariant tensors: the unit matrix $\delta_{ij}$ and the
democratic matrix $D_{ij}$ where all elements are equal to 1. It is
noted that the parameters $y_A^{}$ and $y_A'$ obey the 
relation $y_A^{}=-3y_A'$, which reflects the fact that the tensor
product $3_A\times3_S$ contains only one trivial singlet.

In compactifying the extra dimensions, the boundary conditions on the
bulk scalar fields must be specified. Along the line discussed above,
we impose the following conditions at the two boundaries of the extra
dimension:
\begin{alignat}{2}
  \phi_A(-x^5) &= P_A\,\phi_A(x^5),& \qquad
  \phi_A(x^5+2\pi L_5) &= \phi_A(x^5), 
  \label{twistA}  \\
  \phi_S(-x^6) &= P_S\,\phi_S(x^6),& \qquad
  \phi_S(x^6+2\pi L_6) &= R\,\phi_S(x^6),
\end{alignat}
and all the other boundary conditions are trivial (i.e.,
non-twisted). According to the general recipe, they result in a single
zero mode in each $\phi_A$ and $\phi_S$, and the aligned vacuum is
found to be
\begin{eqnarray}
  \langle\phi_A\rangle \,=\, a\,(0,\,1,-1), \qquad
  \langle\phi_S\rangle \,=\, s\,(1,\,1,\,1).
\end{eqnarray}
The coefficients $a$ and $s$ involve the zero-mode expectation values
and wavefunction factors evaluated at the fixed point where the
standard-model fields reside. Substituting the scalar expectation
values and integrating out heavy modes, we obtain the low-energy
effective model below the compactification scale, which include the
neutrino Yukawa couplings ${\cal L}=\bar\ell_iY_{ij}\nu_jh+\text{h.c}$:
\begin{eqnarray}
  Y \,=\, \begin{pmatrix}
  \,0 & s(y_S^{}+3\bar y_S^{}) & 0 \\
  \,0 & s(y_S^{}+3\bar y_S^{}) & ay_A^{} \\
  \,0 & s(y_S^{}+3\bar y_S^{}) & -ay_A^{} 
  \end{pmatrix}.
  \label{Y}
\end{eqnarray}
Further, the flavor symmetry induces the generation-diagonal form of
right-handed neutrino Majorana 
masses: ${\cal L}_\nu=\overline{\nu^c}M_R\nu$ with $M_R=
\text{diag}(M_1,M_2,M_3)$. Implementing the seesaw mechanism, we
obtain the Majorana mass matrix of light neutrinos
\begin{gather}
  M_L \,=\, m \begin{pmatrix}
  1 & 1 & 1 \\ 1 & 1 & 1 \\ 1 & 1 & 1 \end{pmatrix}
  +m' \begin{pmatrix}
  \,\, & & \\ & 1 & \!\!-1\! \\ & \!-1 & \!\!1 \end{pmatrix}, \\[2mm]
  m\,=\, \frac{s^2(y_S^{}+3\bar y_S^{})^2}{M_2}, \qquad
  m' \,=\, \frac{a^2y_A^2}{M_3}. \quad
\end{gather}
This form of mass matrix is well consistent with the current
experimental results. That is, it leads to the tri-bimaximal generation
mixing~\cite{TBM} and the mass squared 
differences $\Delta m_{21}^2=9|m|^2$ and $\Delta m_{32}^2=
4|m'|^2$, which indicate the normal hierarchy spectrum. A
non-vanishing mass eigenvalue for the first-generation light neutrino
and the deviation from the tri-bimaximal mixing are both generated by
the corrections from higher-dimensional operators 
involving $\nu_1$. The flavor symmetry implies these operators of the
forms $\ell\phi_A^2\nu_1h$ and $\ell\phi_S^2\nu_1h$. Compared with the
leading terms above, they are suppressed 
by $\langle\phi_{A,S}\rangle/\Lambda$ where $\Lambda$ is the cutoff
scale of the theory. Other higher-dimensional operators suppressed by
the cutoff scale are reasonably made small in a similar way. The
corrections to low-energy physics from Kaluza-Klein excited modes
could also be insignificant due to some extra-dimensional
property~\cite{KKsup}. While we do not explicitly discuss the 
charged-lepton sector here, an example of charged-lepton mass matrix
is presented in~\cite{HWY} where the right-handed charged leptons and
a related scalar field are introduced in the four-dimensional
boundary, and the induced hierarchy of charged-lepton mass eigenvalues 
realizes that the total lepton mixing is dominated by the neutrino
tri-bimaximal mixing obtained above. Similar or more reasonable models
might be available in a variety of scenarios which have been discussed
in four or higher-dimensional theory.

The twisting boundary conditions of scalar fields generally induce
flavor-breaking Lagrangian terms on the boundary of 
orbifold. For $\phi_A$, it is easily found that the 
twisting \eqref{twistA} does not affect the induced Yukawa 
couplings $Y$ \eqref{Y} due to the aligned vacuum. On the other hand,
for $\phi_S$, the flavor symmetry breaking could split the matrix
elements in the second column of $Y$. That is ameliorated, for
example, by treating $\phi_S$ as a six-dimensional field (and by
exchanging the two boundary conditions at $x^5=0$ and $\pi R$). In
this case, the fixed point where the standard-model fields live
respects the whole $S_3$ flavor symmetry and so does the induced
Yukawa operators.

\bigskip
\section{$\boldsymbol{A_4}$ flavor symmetry}

The next example is $A_4$ which consists of the even permutations of
four objects. The $A_4$ flavor symmetry has been widely studied in the
literature in view of the recent neutrino experimental
data~\cite{A4}. The $A_4$ group is composed of twelve elements
generated by two fundamental elements $P$, $R$ and their 
derivatives: $PR$, $RP$, $R^2$, $PR^2$, $PRP$, $RPR$, $R^2P$, $RPR^2$,
and $R^2PR$. They satisfy the characteristic relations for 
the $A_4$ group:
\begin{eqnarray}
  P^2 \,=\, (PR)^3 \,=\, R^3 \,=\, 1.
\end{eqnarray}
In addition to the trivial singlet, the $A_4$ group has three
non-trivial representations, the triplet and two pseudo singlets
(denoted as $1'$ and $1''$), the latters of which are respectively
multiplied by the complex numbers $\chi$ and $\chi^2$ under 
the $R$ operation ($\chi=e^{2\pi i/3}$). The representation matrices
for the triplet are built up from
\begin{eqnarray}
  P =\begin{pmatrix}
  \,1 & & \\ & \!-1\! & \\ & & \!-1 \end{pmatrix},
  \qquad\quad
  R =\begin{pmatrix} & \,1\, & \\ & & \,1 \\ \,1 & & \end{pmatrix}.
  \label{RPA4}
\end{eqnarray}
The parity operations $\hat Z$, $\hat Z'$ are written 
by $P$, $RPR^{-1}$ and $R^{-1}PR$. All their representation matrices
have the determinants $+1$. The tensor product of two 
triplets $\phi=(\phi_1,\phi_2,\phi_3)$ and $\varphi=
(\varphi_1,\varphi_2,\varphi_3)$ are given by
\begin{eqnarray}
  \phi\times\varphi &=& 
  (\phi_2\varphi_3,\phi_3\varphi_1,\phi_1\varphi_2)_3
  \,+\,(\phi_3\varphi_2,\phi_1\varphi_3,\phi_2\varphi_1)_3
  \nonumber \\
  && \quad +\,(\phi_1\varphi_1+\phi_2\varphi_2+\phi_3\varphi_3)_1
  \,+\,(\phi_1\varphi_1+\chi^2\phi_2\varphi_2+\chi\phi_3\varphi_3)_{1'}
  \nonumber \\
  && \qquad 
  +\,(\phi_1\varphi_1+\chi\phi_2\varphi_2+\chi^2\phi_3\varphi_3)_{1''},
\end{eqnarray}
where the suffices denote the $A_4$ representations.

\medskip

\subsection{Vacuum alignment}

Let us consider a bulk scalar $\phi$ in the triplet representation 
of $A_4$ symmetry and examine the boundary conditions which leave a
single zero mode in four-dimensional effective theory.

It is first noticed that, if the extra space is one dimensional, only
one type of such boundary conditions is found to be allowed 
for $A_4$ symmetry. This is the Type I boundary condition 
with $T=I$ and $Z\neq I$. In the basis that the parity representation
matrices are flavor diagonal, the other group matrices involve the
rotation $R$ and do not take the form \eqref{WT1}, i.e, they do not
satisfy the consistency relation among the fifth-dimensional
operations. There is no possibility for the Type II boundary condition
since the $A_4$ group matrix elements have the determinants $+1$ which
implies an even number of negative parity signs in each matrix. In
this way, we find only possibility to have a single zero mode 
is $T=I$ and $Z\neq I$. However, with one type of boundary conditions,
we obtain only one type of alignments, which is not compatible with
rich flavor structure of quarks and leptons. This fact leads us to
introduce two spatial extra dimensions for the $A_4$ flavor
symmetry.\footnote{An alternative is to consider additional flavor
symmetry acting on bulk scalar $\phi$. In particular, if $\phi$ is a
negative parity under an additional bulk symmetry, the odd number of
negative parity eigenvalues is possible with boundary conditions
involving this parity twisting. In this case, Type II boundary
condition is also viable within five-dimensional $A_4$ theory.}

\medskip

Unlike the five-dimensional theory, there are several different ways
to compactify the extra two-dimensional space on an orbifold. In this
section, we consider the $T^2/(Z_2\times Z_3)$ ($\simeq T^2/Z_6$)
orbifold. It is however stressed that the result obtained below is also
applicable to $T^2/Z_2$ and $T^2/Z_3$ orbifolds because 
the $Z_2$ and $Z_3$ orbifolding are independent as will be seen. It is
useful to denote the extra-dimensional coordinates $(x^5,x^6)$ by a
complex one $z=x^5+ix^6$. The extra-dimensional space has four types
of operations; the translations $\hat T$ and $\hat T'$, the 
reflection $\hat Z_2$, and the rotation $\hat Z_3$. They act on $z$ as
\begin{alignat}{2}
  \hat T\, &\;:\; z\,\to\,z+2\pi L, & \qquad\qquad
  \hat Z_2 &\;:\; z\,\to\,-z,  \nonumber \\
  \hat T' &\;:\; z\,\to\,z+2\pi L\chi, & \qquad\qquad
  \hat Z_3 &\;:\; z\,\to\,\chi z,
  \label{T2Z6}
\end{alignat}
with $\chi$ being the cubic root of the 
unity ($\chi=e^{2\pi i/3}$). The $T^2/Z_6$ orbifold is obtained by the
identifications with these operations. The $\hat Z_2$ parity has the
four fixed points $z=0$, $\pi L$, $\pi L\chi$ and $\pi L(1+\chi)$. The 
$\hat Z_3$ action leaves the three 
points $z=0$, $(1+2\chi)/3$ and $(2+\chi)/3$ unchanged. Therefore 
the $T^2/Z_6$ orbifold has one fixed point at the origin $z=0$. We
will later consider that the standard model fields live on this
four-dimensional fixed point. On the field 
variables $\phi(x^\mu,z)$, the compactification is performed with the
identification;
\begin{alignat}{2}
  \phi(z+2\pi L) &= T\,\phi(z),& \qquad\qquad
  \phi(-z) &= Z_2\,\phi(z), \nonumber \\[1mm]
  \phi(z+2\pi L\chi) &= T'\phi(z),& \qquad\qquad
  \phi(\chi z) &= Z_3\,\phi(z),
\end{alignat}
where $T$, $T'$ and $Z_{2,3}$ are the triplet representation matrices
of $A_4$ group.

It is important that these matrices should satisfy several consistency
relations associated with the property of orbifold. From the
transformation rules \eqref{T2Z6}, we find the following independent
relations for the $T^2/Z_6$ orbifold:
\begin{gather}
  Z_2Z_3 \,=\, Z_3Z_2, \\
  (Z_2)^2 = (Z_2T)^2 = (Z_2T')^2 \,=\, 1,  \label{T2Z2} \\
  TT'=T'T,  \label{both} \\
  (Z_3)^3 = (Z_3T)^3 \,=\, 1, \qquad
  TZ_3 \,=\, Z_3T'.  \label{T2Z3}
\end{gather}
The first relation means that $Z_2$ and $Z_3$ are commutative and
their representation matrices are simultaneously diagonalized. The
second line implies that, like the $S^1/Z_2$ orbifold, other forms of
parity operations are constructed which are the reflections about 
the $Z_2$ fixed points. The last two types of relations are
characteristic for six-dimensional theory. It is interesting that
these consistency relations rather reduce the number of possible
boundary conditions. We find that only two types of non-trivial
boundary conditions are viable in the $A_4$ theory on $T^2/Z_6$:
\begin{alignat}{2}
  \text{(i)}\,&\;\; Z_2=P,\qquad & Z_3=T=T'=I  \label{A4Z2} \\
  \text{(ii)}&\;\; Z_3=R,\qquad & Z_2=T=T'=I  \label{A4Z3}
\end{alignat}
where the matrices $P$ and $R$ are given in \eqref{RPA4} and $I$ is
the unit matrix. The other sets of matrices with exchanging generation
labels also become the solutions. It is noticed that any non-trivial
boundary condition associated with the translations is not
allowed. The above classification of boundary conditions is general
and valid for any flavor symmetry unless it has non-trivial
commutative matrices $A$ and $B$ which satisfy $A^2=B^3=1$.

For the boundary condition (i), i.e., $\phi(-z)=P\phi(z)$, it is
easily found from the matrix form \eqref{RPA4} that we have a single
zero mode in $\phi_1$ due to a positive parity. The zero mode has a
constant wavefunction profile. Then the vacuum expectation value becomes
\begin{eqnarray}
  \langle\phi\rangle \,\propto\, (1,\,0,\,0).
  \label{A4Z2vac}
\end{eqnarray}
On the other hand, the boundary condition (ii) is explicitly written by
\begin{eqnarray}
  \phi(\chi z) \,=\, R\,\phi(z),
\end{eqnarray}
and the other conditions are trivial and do not affect the physics. It
is convenient to write the matrix $R$ in the diagonal 
form ($\phi=U_R\,\bar\phi$)
\begin{gather}
  U_R^\dagger R\,U_R \,=\, \begin{pmatrix}
  \,1 & & \\ & \chi & \\ && \chi^2
  \end{pmatrix}, \qquad 
  U_R \,=\, \frac{1}{\sqrt{3}}\begin{pmatrix}
  1 & \chi & \chi^2 \\[.5mm]
  1 & \chi^2 & \chi \\[.5mm]
  1 & 1 & 1
  \end{pmatrix}.
  \label{diagR}
\end{gather}
The eigenvalues obtained in \eqref{diagR} indicate that 
only $\bar\phi_1$ has a zero mode with a flat wavefunction. The vacuum
expectation value of the original field $\phi$ thus becomes
\begin{eqnarray}
  \langle\phi\rangle \,\propto\, (1,\,1,\,1).
  \label{A4Z3vac}
\end{eqnarray}
Therefore the alignment of vacuum is established, similarly to 
the $S_3$ flavor theory in five dimensions.

We also have the vacuum alignments on other types of 
orbifolds $T^2/Z_N$ with $N=2,3$ and $4$. For $T^2/Z_2$, the
consistency conditions are a fewer sets than 
for $T^2/Z_6$, i.e.\ \eqref{T2Z2}--\eqref{both}. Therefore in addition
to \eqref{A4Z2} we generally have another solutions for the
consistency conditions and one surviving zero mode. The most general
solutions for $T^2/Z_2$ are given 
by $Z_2, T, T'=\text{1 or $P$}$ (and those with label exchanges). It
is noticed that the aligned vacuum is not affected by this wider
possibility of twisting and is given by \eqref{A4Z2vac}. The same is
true for the $T^2/Z_3$ orbifold. The consistency conditions 
are \eqref{both}--\eqref{T2Z3} and the general solutions for these are
given by $Z_3, T, T'=\text{1 or $R$}$ (and those with changing labels 
and phases). The light-mode physics is not altered with these choices
of boundary conditions and the aligned vacuum 
becomes \eqref{A4Z3vac}. Finally, for the $T^2/Z_4$ orbifold, the
independent consistency conditions 
are $(Z_4)^4=(Z_4^2T)^2=1$ and $TZ_4=Z_4T'$. For the $A_4$ symmetry,
the general solutions are given 
by $Z_4, T, T'=\text{1 or $P$}$ with $T=T'$ (and those with label
exchanges) which leave only one zero mode in low-energy theory. The
vacuum alignment is the same as that on $T^2/Z_2$.

\bigskip
\subsection{Illustrative model}

We apply the above result for the $A_4$ flavor symmetry to
constructing of an explicit orbifold model for lepton masses and
generation mixing. Similarly to the $S_3$ model in the previous
section, a higher-dimensional extra space is introduced for flavor
symmetry invariance, while the vacuum alignment itself is viable in
lower dimensions.

Let us consider an eight-dimensional theory on the 
orbifold $T^2/Z_2\times T^2/Z_3$. The standard-model fields are
assumed to live on a fixed point of the orbifold, which we here choose
the origin of extra-dimensional space. The three families of
left-handed lepton doublets $\ell_i$ transform as a triplet 
of $A_4$ while the right-handed charged 
leptons $e_1$, $e_2$, $e_3$ are assigned to three different 
singlets, $1$, $1'$, $1''$, respectively. We introduce two
gauge-singlet bulk scalars $\phi$ and $\phi'$ of the triplet
representation, which give effective Yukawa and mass operators for
leptons in low-energy theory. The fifth and sixth dimensions are
compactified on $T^2/Z_2$ and the seventh and eighth ones 
on $T^2/Z_3$. The scalars $\phi$ and $\phi'$ live 
on $T^2/Z_2$ and $T^2/Z_3$, respectively.\footnote{The $T^2/Z_2$
orbifold can be replaced with $S^1/Z_2$. In this case, we consider a
seven-dimensional theory compactified 
on $S^1/Z_2\times T^2/Z_3$.} \ The basic building block is the same as
the models presented in~\cite{align2}. The bulk theory has the 
flavor $A_4$ and $Z_3'$ symmetries (other than the 
operation~$\hat Z_3$). The field content and the flavor-symmetry
charges are summarized below:
\begin{eqnarray}
\begin{array}{c|cccccc|cc}
  & \ell & e_1 & e_2 & e_3 & h & \eta & \phi & \phi' \\ \hline
A_4 & 3 & 1 & 1' & 1'' & 1 & 1 & 3 & 3 \\
Z & \chi & \chi & \chi & \chi & 1 & \chi & \chi & 1
\end{array}
\end{eqnarray}
Here $h$ and $\eta$ are the electroweak doublet and singlet Higgs
fields, respectively. We have boundary interactions between the bulk
and boundary fields which are invariant under the flavor symmetry:
\begin{eqnarray}
  {\cal L}_Y \,=\, y_1\bar e_1 \ell\phi h 
  +y_2\bar e_2\phi'\ell h  +y_3\bar e_3\ell\phi' h
  +w_1\phi\overline{\ell^c}\ell h^2 +w_2\eta\overline{\ell^c}\ell h^2
  +\cdots,
  \label{LY_A4}
\end{eqnarray}
where $w_{1,2}$ and $y_{1,2,3}$ are the coupling constants and the
wavefunctions of bulk scalar fields are evaluated at the fixed point
on which the standard-model fields reside.\footnote{For the mode
expansion on the $T^2/Z_3$ orbifold, see for example Ref.~\cite{T2Z3}.}

To fix the model, the boundary conditions for bulk fields must be
specified. According to the result obtained in the above, we have
the following non-trivial boundary conditions on the scalar fields: 
\begin{eqnarray}
  \phi(-z) &=& P\,\phi(z),  \label{illustZ2} \\
  \phi'(\chi z') &=& R\,\phi'(z'),  \label{illustZ3}
\end{eqnarray}
where $z=x^5+ix^6$ and $z'=x^7+ix^8$, and the other boundary
conditions are trivial. Under these conditions, the previous analysis
says that the zero modes are found 
in $\phi_1$ and $\phi'_1+\phi'_2+\phi'_3$ which develop the
expectation values of the form
\begin{eqnarray}
  \langle\phi\rangle \,=\, a\,(1,\,0,\,0), \qquad
  \langle\phi'\rangle \,=\, a'(1,\,1,\,1),
\end{eqnarray}
in low-energy effective theory below the compactification scale. It is
noticed that these aligned forms of expectation values were required
in the models of Ref.~\cite{align2}. Inserting the scalar expectation
values in \eqref{LY_A4}, we obtain the operators for charged-lepton
Dirac masses and neutrino Majorana masses; ${\cal L}_M=
\bar e_i(M_e)_{ij}\ell_j+\overline{\ell^c_i}(M_\nu)_{ij}\ell_j$:
\begin{eqnarray}
  M_e \,=\, a'v\!\begin{pmatrix}
  y_1 & y_1 & y_1 \\
  y_2 & \chi^2 y_2 & \chi y_2 \\
  y_3 & \chi y_3 & \chi^2 y_3
  \end{pmatrix}, \qquad
  M_L \,=\, v^2\!
  \begin{pmatrix}
  w_2b & & \\
  & \!\!w_2b & \!w_1a \\
  & \!\!w_1a & \!w_2b
  \end{pmatrix},
\end{eqnarray}
where $v=\langle h\rangle$ and $b=\langle\eta\rangle$. These forms of
mass matrices are known~\cite{align2} to be well fitted to the present
experimental values. The lepton generation mixing is exactly given by
the tri-bimaximal mixing. The neutrino mass eigenvalues are predicted
as $\Delta m^2_{21}=-\text{Re}[w_1a(w_1a+2w_2b)]v^2/2$ and 
$\Delta m^2_{31}=-4\text{Re}(w_1w_2ab)v^2$. The hierarchy of
charged-lepton masses is proportional to that of $y_i$. It may be
implemented by, for example, the Froggatt-Nielsen 
mechanism~\cite{FN} or extra-dimensional scheme~\cite{ExDsup}, where
the three right-handed charged leptons have different properties.

Finally, we comment on the boundary 
conditions \eqref{illustZ2}--\eqref{illustZ3} and the symmetry
invariance of the boundary Lagrangian ${\cal L}_Y$. As mentioned
before, the consistent boundary conditions 
for $T^2/Z_2$ and $T^2/Z_3$ have wider possibilities 
than \eqref{A4Z2} and \eqref{A4Z3}. An important point is that the
vacuum alignment itself is not affected by such choices of twisting as
long as the light-mode physics is concerned. This fact can, in turn,
be utilized to protect boundary terms by the flavor symmetry. Namely,
flavor symmetry breaking occurs within the hidden sector which is
separate in the extra-dimensional space from the visible sector where
the standard-model fields reside. Such boundary conditions are given,
for example, by $Z_2=1$ and $T=T'=P$ for $T^2/Z_2$, and $Z_3=1$ and 
$T=T'=R$ for $T^2/Z_3$. Bulk scalar fields interact both of these
sectors and induce effective mass operators by connecting up the
symmetry-invariant terms on the visible boundary and the aligned
vacuum expectation values on the hidden boundary.

\bigskip
\section{Summary}

In this paper we have presented the scenario for breaking flavor
symmetry with Scherk-Schwarz twisted boundary conditions on bulk
scalar fields, where only one zero mode from multiple scalar fields
survives in low-energy effective theory. For non-abelian discrete
flavor symmetry, the vacuum alignment is shown to be achieved, that
is, the symmetry-breaking expectation values of these scalars are
aligned to definite directions. This implies that flavor symmetry in
the high-energy regime can leave some imprints to the low-energy
regime and provides a viable explanation for the fermion mass
problems. Our scheme of vacuum alignment needs no elaboration of
analyzing complicated scalar potential which generally involves
multiple scalar fields and requires extra symmetries to realize the
vacuum alignment.

As applications of our approach, the $S_3$ and $A_4$ flavor symmetries
are considered in five and six dimensions compactified on 
the $S^1/Z_2$ and $T^2/Z_N$ ($N=2,3,4,6$) orbifolds, and possible
types of boundary conditions are classified. Further we have
constructed explicit models for lepton masses and mixing, and shown 
that simpler setups lead to realistic phenomenology such as the
tri-bimaximal generation mixing of neutrinos. Other possibilities
including the quarks sector/grand unification, curved gravitational
backgrounds, geometrical origins of flavor symmetries in extra
dimensions~\cite{origin} and so on will be investigated in future study.

\bigskip
\subsection*{Acknowledgments}
\noindent
This work is supported by the scientific grants from the ministry of
education, science, sports, and culture of Japan No.~20540266 and
20740135) and also by the grant-in-aid for the global COE 
program "The next generation of physics, spun from universality and
emergence" and the grant-in-aid for the scientific research on
priority area (\#441) "Progress in elementary particle physics of the
21st century through discoveries of Higgs boson and 
supersymmetry" (No.~16081209). The work of Y.O.\ is supported by the
Japan society of promotion of science (No.~20$\,\cdot\,$324).


\end{document}